\documentclass[11pt]{article}
\usepackage{graphicx,floatflt,amssymb,epsf}
\usepackage{epsfig}
\def\n1{{\cal{N}}_{1i}} 
\newcommand{\ba}{\begin{array}}    
\newcommand{\ea}{\end{array}}    
\newcommand{\bd}{\begin{displaymath}}    
\newcommand{\ed}{\end{displaymath}}    
\newcommand{\be}{\begin{equation}}    
\newcommand{\ee}{\end{equation}}    
\newcommand{\bea}{\begin{eqnarray}}    
\newcommand{\eea}{\end{eqnarray}}    
\newcommand{\email}{E-mail address:}
\begin{document} 
\begin{flushright} 
HRI-P 08-09-002\\ 
RECAPP-HRI-2008-013
\end{flushright}
\vskip 5pt 

\begin{center}
{\Large{\bf The impossibility of heavy neutrino dark matter in the Littlest 
Higgs Model with T-parity: constraints from direct search }}
\vskip 25pt {\bf Paramita Dey$^\star$\footnote{\tt\email
paramita@mri.ernet.in}}, {\bf Sudhir Kumar Gupta$^\dagger$\footnote{\tt\email
skgupta@iastate.edu}} and {\bf Biswarup
Mukhopadhyaya$^\star$\footnote{\tt\email biswarup@mri.ernet.in}}
\vskip 10pt
{\em $^\star$\hspace{-.2cm} 
Regional Centre for Accelerator-based Particle Physics \\
Harish-Chandra Research Institute \\
Chhatnag Road, Jhusi, Allahabad--211019, India\\[.2cm]

$^\dagger$\hspace{-.2cm} 
Department of Physics \& Astronomy \\ 
Iowa State University\\
Ames, IA--50011, USA}\\[.2cm]

\normalsize 
\end{center} 
\begin{abstract} 

  We consider the Littlest Higgs Model with T-parity (LHT), in the parameter
  region where a heavy neutrino is the lightest T-odd particle (LTP). Having
  emphasized that this corresponds to a sizable region in the parameter space
  of the theory, we show that both the Cryogenic Dark Matter Search (CDMS) and
  Xenon10 experiments disallow the entire region where the masses of the new
  particles in LHT can lie within several TeV. Therefore, any observation of
  the signals of a heavy neutrino LTP is likely to seriously reopen the issue
  of cold dark matter in the universe.

\end{abstract} 
\vskip 30pt

\setcounter{footnote}{0} 
\renewcommand{\thefootnote}{\arabic{footnote}} 

{\bf Introduction}: In the last few years the existence of a dark matter (DM)
candidate, comprising about $23\%$ of the energy density of the Universe, has
been firmly established by cosmological observations, of which the WMAP
\cite{wmap} results are most recent and notable. Studies on the large scale
distribution of galaxies as well as the anisotropy of the cosmic microwave
background radiation (CMBR) disfavor hot dark matter as the primary DM
component. However, the exact nature of cold dark matter (CDM) is largely
unknown, and a vigorous experimental effort is devoted to the explication of
its nature. If CDM is of particle physics origin, then one is forced to
postulate a new elementary weakly interacting massive particle (WIMP), which
must be stable.

Such particles occur naturally in several extensions of the standard model
(SM). A typical CDM candidate can be a Dirac or a Majorana fermion, a vector
boson or a scalar. Its mass may range anywhere from a few GeV to a few TeV.
Rather interesting implications are thus suggested for collider experiments
and in direct searches via elastic scattering on target nuclei. Its footprints
are also expected in astrophysical observations such as gamma ray bursts from
galactic centers. Artifacts of dark matter annihilation in the galactic halo
or the center of the sun are also objects of recent investigation.

The new physics theories which can accommodate a CDM candidate must also
provide an explanation of its stability (though of course, one may assume
something {\em ad hoc}, such as heavy stable fourth generation neutrinos
\cite{khlopov}). This is done in a large class of models through a $Z_2$
symmetry against which the candidate particle is odd, with no other $Z_2$-odd
particle below it in the spectrum. This happens in supersymmetric (SUSY)
models \cite{susy1}, theories with universal extra dimensions (UED)
\cite{ued,tait} as well as little Higgs models with T-parity. We focus on the
last of these scenarios in this note.

Little Higgs theories \cite{lhm1,lhm2} form a class of models where the Higgs
mass is stabilized via a new physics $f$ ($\sim$ TeV) at which the breakdown
of a global symmetry gives rise to the standard model Higgs boson and host of
other scalar as Goldstone bosons. The Higgs mass is generated by the
Coleman-Weinberg mechanism. However, $f\lesssim$ TeV is not found to be easily
compatible with precision electroweak constraints, and some additional
postulates are necessary.

Tree-level violation of precision constraints is avoidable through a discrete
symmetry called T-parity is, for example, the Littlest Higgs Model [LHT]
\cite{lhtp,lhew}. All the particles in such a spectrum can be classified as
$T$-even/odd, and the lightest $T$-odd particle (LTP) turns out to be a CDM
candidate. Over a large part of the parameter space of this model, the LTP is
a spin-1 particle (the heavy photon or $A_H$) whose implications as a CDM have
been studied extensively \cite{lhtdm,lhtdm_exp,barger}.  However, a spin-$1/2$
neutral Dirac fermion (the heavy neutrino or $\nu_H$) becomes lighter than the
heavy photon over a certain region which is otherwise viable
phenomenologically. Thus the $\nu_H$ becomes the LTP in this region. Since
this region in the parameter space is phenomenologically distinct from that
with a heavy photon LTP, it is important to make a clear statement on whether
this region is allowed by the extant results on direct dark matter searches.
Here we probe this territory of the LHT model, and study the possibilities of
this heavy neutrino LTP in direct detection experiments
\cite{cdms,xenon,xenon10}.

{\bf The LTP of the LHT model}: In the Littlest Higgs model, a global $SU(5)$
spontaneously breaks down to $SO(5)$ at a scale $\Lambda~=~4\pi f$, with $f
\simeq 1~{\rm TeV}$. An $[SU(2)\otimes U(1)]^2$ gauge symmetry is imposed.
This gauge group breaks simultaneously into the diagonal subgroup
$SU(2)_L\otimes U(1)_Y$, which is identified as the SM gauge group. One thus
has four heavy gauge bosons $W^\pm_H$, $Z_H$ and $A_H$ with masses $\sim f$,
in addition to the SM gauge fields. The SM Higgs doublet $ H$ is part of an
assortment of pseudo-Goldstone bosons, together with a heavy $SU(2)$ triplet
scalar $\Phi$, resulting from the spontaneous breaking of the global symmetry.
The augmented symmetry controls quadratically divergent contributions to the
Higgs mass. Finally, the Coleman-Weinberg mechanism leads to a radiatively
generated Higgs mass which naturally remains within a TeV. The input used for
making relatively low values of $f$ consistent with all precision electroweak
observables is a discrete symmetry called $T$-parity, which maps the two pairs
of gauge groups $SU(2)_i\otimes U(1)_i$, $i=1,2$ into each other, forcing the
corresponding gauge couplings to be equal. All SM particles are even under
$T$-parity, while the four additional massive gauge bosons and the Higgs
triplet are $T$-odd. In order to render the fermionic sector consistent with
T-parity and gauge invariant at the same time, one has to introduce additional
heavy vector-like fermions for each family.  Particular linear combinations
of the fermions transforming under each of the two SU(2)'s yield the SM quarks
and leptons, while the orthogonal combinations give us T-odd heavy fermions
$\{u^i_H, d^i_H\}$ and $\{l^i_H, \nu^i_H\}$ for $i=1,2,3$, which are
vector-like doublets under the SM SU(2).  The requirement of cancellation of
quadratic divergence in the Higgs mass further prompts one to postulate two
extra heavy fermionic partners for the top quark, one of which is $T$-even and
the other $T$-odd (see \cite{lhtp} for details).  The multiplicative
conservation of T-parity prevents the lightest T-odd state from further
decays, thus making it the LTP and the CDM candidate.

\begin{figure} 
  \vspace{-10pt}
  \centerline{\hspace{-3.3mm} \rotatebox{0}{\epsfxsize=8cm\epsfbox{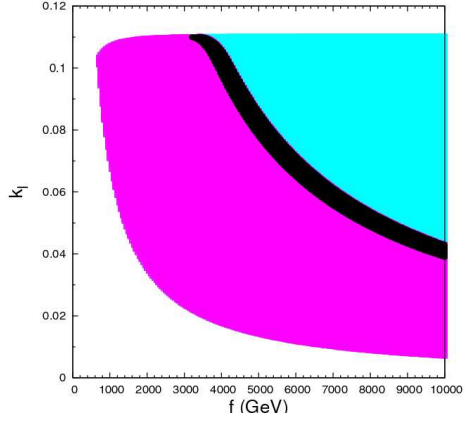}}}
  \hspace{3.3cm}\caption{\small\sf The region of the parameter space of the
    LHT model in the $\kappa_l-f$ plane (colored region) corresponding to
    $\nu_H$ as the LTP. The black band is strictly allowed by the WMAP
    observation ($\Omega_{\mbox{\tiny DM}} h^2 = 0.105^{+0.007}_{-0.013} $).
    In the pink (dark grey) region, there is a shortfall in the contribution
    to the relic density, while the blue (light grey) region corresponds to
    excessive relic density. The region corresponding to $m_{l_H}<100~{\rm
    TeV}$, disallowed by the LEP experiment, has been excluded from the
    colored patches. As will be seen from the text, the entire colored region
    is disallowed from direct search for dark matter.}  \protect\label{ps}
\end{figure}

The masses of the heavy gauge bosons are dictated by the scale $f$ which can
be as low as $500~{\rm GeV}$
\cite{Hubisz_et_al,Asano_et_al,Hundi_Mukhopadhyaya_Nyffeler,constraints_flavor_physics},
while the masses of the heavy leptons (quarks) are additionally determined by
a parameter $\kappa_l$ ($\kappa_q$)\footnote{In principle, $\kappa$ can be a
  $3\times 3$ matrix carrying flavor indices, i.e.\ $m^{ij}_{l_h,q_h}\sim
  \kappa^{ij}_{l_H,q_H} f$. We have simplified our analysis by assuming
  $\kappa^{ij}_{l_H} = \kappa_l \delta^{ij}$.}, where $\kappa\le 4.8$ (for
$f\sim 1~{\rm TeV}$) \cite{Hubisz_et_al}. In particular, the masses of the
heavy photon, the heavy neutrino and the heavy charged lepton are given by
\begin{eqnarray}
\label{masses}
m_{A_H} &=& \frac{fg'}{\sqrt 5} \left( 1-\frac{5v^2}{8f^2} \right),
~~~~~ m_{Z_H} = {fg} \left( 1-\frac{v^2}{8f^2} \right),  \nonumber \\
m_{nu_H} &=& \sqrt 2\kappa_l f \left( 1-\frac{v^2}{8f^2} \right), ~~~ m_{l_H}
= \sqrt 2 \kappa_l f.
\end{eqnarray}
This clearly indicates that small values of $\kappa_l$ will lead to
$m_{\nu_H}<m_{A_H}$ making $\nu_H$ the LTP; otherwise $A_H$ plays that role.
The colored region of Figure \ref{ps} shows the region on the $\kappa_l-f$
plane corresponding to $\nu_H$ LTP. The constraints from the production of
$l_H$-pairs at the Large Electron Positron (LEP) experiment has been taken
into account in marking the allowed region. Note that for every $f$ there
exists one maximum and one minimum value of $\kappa_l$. The maximum $\kappa_l$
is determined from the requirement $m_{\nu_H}<m_{A_H}$. For large $f$, this
translates into $\kappa_l<fg'/\sqrt{10}$ (neglecting corrections $\sim
v^2/f^2$ in Eq.\ (\ref{masses})), and the upper limit therefore becomes almost
independent of $f$. The lower limit on $\kappa_l$ is set by the fact that
$m_{\nu_H}>m_{Z}/2$ so that the $Z$ does not decay into a pair of $\nu_H$.
Thus for large $f$, under the same approximations as before, the minimum
$\kappa_l$ is almost independent of $f$. For smaller values of $f$ however, as
the factor $f\left( 1-{v^2}/{8f^2} \right)$ becomes smaller, the minimum of
$\kappa_l$ rapidly grows to larger values to maintain $m_{\nu_H}>m_{Z}/2$.
Note that the heavy electron $e_H$, becomes almost degenerate with $\nu_H$ for
large values of $f$. In Figure \ref{ps} we have also ensured that the mass of
${l_H}$ is more than that of ${\nu_H}$ by $\sim 0.51 ~{\rm MeV}$, so that
$e_H$ does not become stable on the cosmological scale.

Thus within the colored region enclosed by the curve in Figure \ref{ps}, the
possible CDM candidate from the LHT model is the heavy Dirac neutrino
${\nu_H}$ and not $A_H$ (the latter corresponds to the white region in Figure
\ref{ps}). While the viability of $A_H$ from direct dark matter search and the
associated phenomenology have been studied in detail
\cite{lhtdm_exp,barger,more}, we extend this study to a $\nu_H$ CDM. Though
the observations pertaining to theories such as UED are broadly valid here,
our emphasis is on the part of the LHT parameter space constrained in this
manner. The colored region in Figure \ref{ps} is further constrained from
relic density bounds. The region marked in green corresponds to relic density
below 0.092, and thus leads to underclosure (in which case it cannot account
for all dark matter but can still be a viable candidate). The region marked in
red corresponds to relic density above 0.112, being thus by and large
disallowed by the WMAP results due to overclosure. The blue-colored band,
corresponding to relic density between 0.112 and 0.092, is the WMAP-allowed
region where the ${\nu_H}$ is the lone CDM candidate. The relic density
calculation for all the cases has been done using the package micrOMEGAs 2.2.

It should be noted here that, if one has to {\it exactly} fit the WMAP data
with $\nu_H$ dark matter, then one requires a minimum value of $f$ on the
order of $3~{\rm TeV}$. This, however, leads to such values of the $W_H$,
$Z_H$ masses, which tend to shift the Higgs mass to well above a TeV, thus
requiring some fine-tuning, and introducing a `little hierarchy'. From this
angle, a $\nu_H$ LTP is somewhat disallowed theoretically, especially if one
uses it to account for all dark matter.

{\bf Signature of $\nu_H$ dark matter}: Typically, the direct detection of a
WIMP involves elastic scattering of the WIMP with a nucleus in a detector. The
nucleus recoils with some energy whose distribution is a function of the
masses of the WIMP and the nucleus, Thus the WIMP-nucleon (average)
cross-section calculated in a specific model is the starting point, and a
region in the parameter space is ruled out if the prediction exceeds the upper
limit obtained from the absence of recoil events satisfying the appropriate
cuts.

The Cryogenic Dark Matter Search (CDMS) experiment \cite{cdms}, for example,
is designed to detect atomic nuclei in germanium (Ge) and silicon (Si)
crystals that have been scattered by the incident WIMPs, while XENON10
\cite{xenon,xenon10} uses liquid Xenon (Xe) as a sensitive detector medium.
Generically the WIMP-nucleus interactions can be split into spin-independent
(scalar) and spin-dependent parts. The scalar interactions add coherently in
the nucleus, so that the heavier the nuclei the better is the sensitivity.
Spin-dependent interaction, on the other hand, relies mainly on one unpaired
nucleon, and thus dominates over scalar interactions for light nuclei. On the
whole, the cross-section for the WIMP-nucleus interaction is typically low, so
that large detectors are required.

\begin{figure} 
  \vspace{-10pt}
  \centerline{\hspace{-3.3mm} \rotatebox{0}{\epsfxsize=8cm\epsfbox{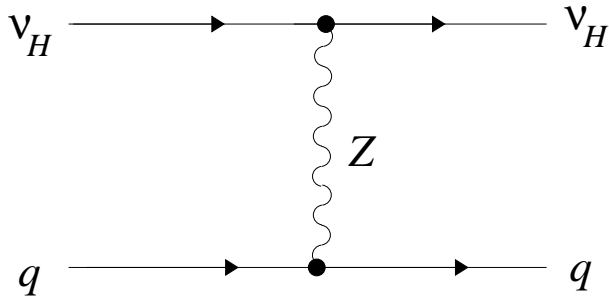}}}
  \hspace{3.3cm}\caption{\small\sf Leading order Feynman graph for effective
    $\nu_H$-quark scattering through the exchange of the SM $Z$-boson.}
  \protect\label{fg}
\end{figure}

To arrive at the WIMP-nucleus cross-section, one has to start with
interactions at the quark level. For the case in study the leading
contribution is shown in Figure \ref{fg} where the $\bar\nu_H \nu_H Z$
interaction is given by
\begin{equation}
{\cal L} = \frac{g}{2\cos\theta_W}\bar\nu_H \gamma_\mu \nu_H Z^\mu 
\label{vertex}
\end{equation}
The coupling of $\nu_H$ with $W$, $Z$ is vector-like \cite{Hubisz_Meade}.
Using this the WIMP-quark matrix element can be computed, and then it has to
be converted into effective couplings of the WIMP to protons and neutrons
\cite{dm_review,micro}, namely $\lambda_p$ and $\lambda_n$. This is an
effective vector-vector four-fermion interaction, for which the
spin-independent cross-section dominates. The effective couplings $\lambda_p$
and $\lambda_n$ are given as
\begin{equation}
\lambda_p = 2\lambda_u + \lambda_d = \frac{e^2}{4\sin^2\theta_WM^2_W} ~ \left[
\frac{1}{2} (1-4\sin^2\theta_W) \right], ~~ \lambda_n = 2\lambda_d + \lambda_u
= -\frac{e^2}{4\sin^2\theta_WM^2_W} ~ \left[ \frac{1}{2} \right],
\end{equation}
where $\lambda_{u,d} = \frac{e^2}{4\sin^2\theta_WM^2_W} (T^{u,d}_3 - 2Q_{u,d}
\sin^2\theta_W)$ are the strengths of WIMP-quark interactions. Starting from
the input Lagrangian shown in Eq.\ (\ref{vertex}), the WIMP-nucleon
cross-section can be computed following a procedure similar to that in
\cite{dm_review}. This yields
\begin{equation}
\sigma^{\rm SI}_0 =
\frac{4\mu^2_{\nu_H}}{\pi} \left[\lambda_pZ+\lambda_n(A-Z)\right],
\label{sigma}
\end{equation}
where $\mu_{\nu_H} = M_{\nu_H} M_Z / (M_{\nu_H} + M_Z)$ is WIMP-nucleon
reduced mass, $Z$ is the number of protons and $(A-Z)$ is the number of
neutrons in the detector nucleus. Note that $\sigma^{\rm SI}_0$ is the
cross-section for the WIMP scattering at rest from a point-like nucleus being
known as the `standard' cross-section at zero momentum transfer. To obtain the
cross-section precisely, one has to convolute $\sigma^{\rm SI}_0$ with the
nuclear form factor $F(Q)$ where $Q$ is the energy transferred from the WIMP
to the nucleus, and then integrate over $Q$. However, for an order of
magnitude estimation of the WIMP-nucleus cross-section, estimation of
$\sigma^{\rm SI}_0$ alone is sufficient \footnote{Our calculation has been
  cross-checked against results using the package micrOMEGAs 2.2
  \cite{micro,Belanger:2006is}.}. This is because the energy exchange between
the WIMP and the nucleus is on the order of a few hundreds of KeV. Given such
small recoil energy compared to nucleon masses, the inclusion of the nuclear
form-factor is not expected to yield cross-sections drastically different from
$\sigma^{\rm SI}_0$. Finally, the scattering cross-section per nucleon is,
\begin{equation}
\sigma^{\rm SI}_{nuc} = \sigma^{\rm SI}_{0} \frac{m^2_{nuc}}{\mu^2_{\nu_H}
A^2}.
\label{sigmanuc}
\end{equation}

\begin{figure} 
  \vspace{-10pt}
  \centerline{\hspace{-3.3mm} \rotatebox{0}{\epsfxsize=8cm\epsfbox{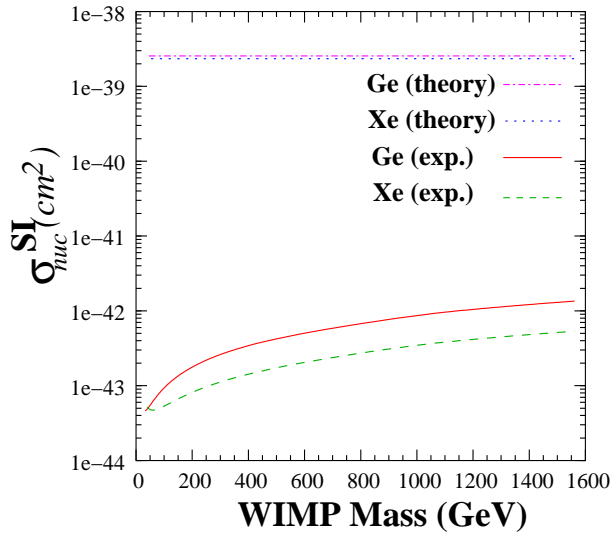}}}
  \hspace{3.3cm}\caption{\small\sf Theoretical estimates of spin-independent
    cross-sections for $\nu_H$-nucleon scattering, for the CDMS and Xenon10
    experiments. The corresponding experimental limits on the WIMP-nucleon
    cross-section are also shown, clearly indicating that the entire colored
    region in Figure 1 is disallowed.}  \protect\label{excl}
\end{figure}

The scattering cross-section per nucleon estimated from Eq.\ (\ref{sigmanuc})
is plotted in Figure \ref{excl} as a function of the WIMP mass. Since
$\sigma^{\rm SI}_{nuc}$ is independent of the WIMP mass, the plot is a
straight line parallel to the $x$-axis with an estimated value $\sigma^{\rm
  SI}_{nuc} \sim 2.34\times 10^{-39}{\rm cm}^2$ for Ge and $2.55\times
10^{-39}{\rm cm}^2$ for Xe respectively. Figure \ref{excl} also shows the
existing experimental limits on the cross-section from CDMS \cite{cdms} and
Xenon10 \cite{xenon} are drawn by the red and the green lines respectively.
Note that the experimental plots have been linearly extrapolated to a
WIMP-mass $\sim 1.6$ TeV, a value which corresponds to $f=10$ TeV (and
$\kappa=0.1$). For larger values of $f$, the little hierarchy problem crops up
in the LHT-model, and it becomes phenomenologically uninteresting. Figure
\ref{excl} clearly shows that the theoretical estimation of $\sigma^{\rm
  SI}_{nuc}$ is way beyond the limits obtained from the experiments. {\em Thus
  a $\nu_H$ LTP, in the entire colored region in Figure 1, is ruled out, at
  least upto $f=10$ TeV. An even further extrapolation does not leave any room
  for a $\nu_H$ CDM, unless one goes way above the $10$ TeV mark in $f$,
  something that cannot be motivated from the stabilization of the electroweak
  scale}.

Is this study, focusing on the $f - \kappa_l$ plane, unduly restricted?  The
answer is no, for the following reasons. First of all, the LTP can be one of
three particles, namely, $Z_H$, $A_H$ and $\nu_H$. As is obvious from Eq.\
(\ref{masses}), one always has $m_{Z_H} > m_{A_H}$. This leaves us with two
possibilities only, of which one, namely an $A_H$ LTP, has been studied
extensively. We take up the remaining one here, and establish its
impossibility.  We emphasize that our demonstration is {\em not affected} on
varying the rest of the parameters of the model. Over the region with a
$\nu_H$ LTP, $\kappa_q$ must be such as to make the heavy quark more massive
than the heavy leptons, but there is no further dependence on its value in the
cross-section pertinent to dark matter detection. In a similar vein, any
departure from $\kappa_q^{ij} = \delta^{ij}$, on which limits are imposed from
various flavor-changing processes \cite{constraints_flavor_physics}, does not
affect our conclusions.  Non-diagonality as well as non-universality in
$\kappa_l$ is even more restricted from the limits on lepton flavor violating
phenomena \cite{constraints_flavor_physics}. Moreover, such non-universality
does not prevent a $\nu_H$ from being the LTP, in which case our Figure 1 will
contain that particular $\kappa_l$ to which its mass is related. A heavy
neutrino LTP is completely disallowed in such a situation as well.

{\bf Conclusion}: In summary, we probe the particular parameter region in the
$\kappa_l-f$ plane of the Littlest Higgs model, which corresponds to the heavy
neutrino $\nu_H$ as the lightest $T$-odd particle. We then estimate the
spin-independent scattering cross-section for $\nu_H$ with the Germanium and
Xenon nuclei, and compare them with the limits obtained from the CDMS and
Xenon10 experiments. We find that the possibility of $\nu_H$ being the WIMP is
ruled out upto very large values of $f$ ($>> 10$ TeV). Although it may not be
straightforward to identify a $\nu_H$ LTP at a hadron collider, a careful
analysis of (leptons + $E_T{\!\!\!\!\!\!/\ }$) final states at, say, a linear
electron-positron collider may supply crucial information on its identity. Our
study serves to establish that direct dark matter searches forbid such a final
state in the LHT scenario, unless $f$ is so large that the model itself
becomes phenomenologically irrelevant.

{\bf Acknowledgment}: We thank A. Bandyopadhyay and S. Raychaudhuri for
technical help and U. Chattopadhyay for helpful discussions. This work was
partially supported by funding available from the Department of Atomic Energy,
Government of India for the Regional Centre for Accelerator-based Particle
Physics, Harish-Chandra Research Institute.


\end{document}